\documentclass[aps,pre,floatfix,superscriptaddress,twocolumn,10pt]{revtex4-1}

\usepackage{graphicx}% Include figure files
\usepackage{dcolumn}% Align table columns on decimal point
\usepackage{bm}% bold math
\usepackage[dvipsnames]{xcolor}
\usepackage{hyperref}% add hypertext capabilities
\usepackage{epsfig}
\usepackage{comment}
\begin{document}

\title{Entropy of full covering of the kagome lattice by straight trimers}

\author{Deepak Dhar}
\email{deepakdhar1951@gmail.com}
\affiliation{International Center for Theoretical Sciences, Tata Institute of Fundamental Research, Bengaluru, 560089, India}
\author{Tiago J. Oliveira}
\email{tiago@ufv.br}
\affiliation{Departamento de F\'isica, Universidade Federal de Vi\c cosa, 36570-900, Vi\c cosa, Minas Gerais, Brazil}
\author{R. Rajesh} 
\email{rrajesh@imsc.res.in}
\affiliation{The Institute of Mathematical Sciences, C.I.T. Campus, Taramani, Chennai 600113, India}
\affiliation{Homi Bhabha National Institute, Training School Complex, Anushakti Nagar, Mumbai 400094, India}
\author{J\"urgen F. Stilck}
\email{jstilck@id.uff.br}
\affiliation{Instituto de F\'isica and National Institute of Science and Technology for Complex Systems,
Universidade Federal Fluminense, Niter\'oi, RJ, Brazil}
\date{\today}

\begin{abstract}

We consider the number of ways all the sites of a kagome lattice can be covered by non-overlapping linear rigid rods where each rod covers 3 sites. We establish a 2-to-1 correspondence between the configurations of trimers on the kagome lattice to the covering by dimers of a related hexagonal lattice to show that entropy of coverings per trimer $s_{\text{tri,kag}}$ equals the entropy per dimer $ s_{\text{dim,hex}} $, and is given by $ s_{\text{tri,kag}} = s_{\text{dim,hex}} = \frac{1}{2 \pi} \int_0^{ 2 \pi/3} \log( 2 + 2 \cos k) dk \approx 0.323065947\ldots$.

\end{abstract}

\maketitle

%\tableofcontents

\section{Introduction}

The macroscopic properties of hard-core systems of rod-like molecules have been studied extensively since Onsager’s seminal result~\cite{onsager1949effects} that long hard cylinders in three dimensions undergo an isotropic–nematic transition at sufficiently high density. Subsequent work in this area has explored different shaped molecules: spheres, triangles, rectangles, ellipsoids, etc.~\cite{frenkel1988structure,donev2004jamming,verberkmoes1999triangular,dhar2023phase}. The study of lattice models of hard-rod systems was initiated by Flory and Zwanzig~\cite{flory1956phase,zwanzig1963first}. The simplest such models consists of straight $k$-mers, which occupy $k$ consecutive sites on a lattice. The study of these lattice models can be classified into two categories: the limit of full packing where all the sites of the lattice are covered with rods and entropy and correlations between rods being of interest; and away from full packing when vacancies are present and different phases and phase transitions among them being of interest.

The full-packing  problem in two dimensions is of special interest. In this case, the tilings admit a vector height-field representation~\cite{kenyon2000conformal}, giving rise to nontrivial orientational correlations driven purely by geometric constraints. 
The special case of dimers ($k=2$) is a very well studied problem, for which there is a vast literature and many exact results~\cite{kasteleyn1961statistics, kasteleyn1963dimer, temperley1961dimer, fisher1961statistical,  fisher1963statistical, kenyon2003introduction, nagle1989dimer, wu2006dimers, huse2003coulomb}. For dimers, the entropy per site on the square lattice is exactly solvable and equals $G/\pi=0.29156\ldots$, where $G$ is Catalan’s constant~\cite{kasteleyn1961statistics}. On the hexagonal lattice the entropy per dimer is known to be $0.323065947\ldots$. We note that this result for the honeycomb lattice  was determined  in 1950 by Wannier using the equivalence of the problem to finding the zero-temperature entropy per site of the antiferromagnetic Ising model on the triangular lattice~\cite{wannier1950antiferromagnetism}, before the work of Kasteleyn~\cite{kasteleyn1961statistics, kasteleyn1963dimer} and Fisher~\cite{fisher1961statistical,  fisher1963statistical}.  For $k=3$ or trimers, numerical diagonalization of transfer matrices for finite strips yields an entropy per site of $0.158520 \pm 0.000015$ in two dimensions~\cite{ghosh2007random}.
This value has been numerically confirmed in more recent works~\cite{pasinetti2021entropy, rodrigues2022entropy, rodrigues2023entropy}, where estimates were obtained for the entropy on the square lattice, $s_{square}$, for $k$ up $10$.

In higher dimensions, results for the full packing are fewer.  In three dimensions, exact solutions for the entropy of dimers are known for certain noncubic lattices, where correlations are strictly finite-ranged~\cite{2008-dc-prl-exact}.  For all $d\geq 2$,  for large $k$, the entropy per site  was shown to be exactly $ k^{-2} \log k$ to leading order in $k$ ~\cite{dhar2021entropy}.  

When vacancies are present, interesting phases arise. In two dimensions, for $k\geq 7$, Monte Carlo simulations reveal the existence of a continuous isotropic-nematic transition for large enough $k$ as the rod density is increased \cite{ghosh2007orientational, matoz2008determination, matoz2008critical, matoz2008critical1}.  A second transition has been observed at higher densities, from nematic to a phase without orientational order~\cite{2012-krds-aipcp-monte,2013-krds-pre-nematic}. In the limit of large $k$, it has been argued that this second phase transition is discontinuous~\cite{shah2022phase}.

In three dimensions, from Monte Carlo simulations, it is found that  $k \geq 7$ exhibit a sequence of phase transitions—from a disordered phase to a nematic phase, and subsequently to a layered disordered phase—as the density increases. For $4<k<7$, the nematic phase is absent, and the system instead undergoes a single transition directly from a disordered to a layered disordered phase~\cite{vigneshwar2017different, gschwind2017isotropic}. The behavior of the transition has also been studied on the Bethe lattice, and other locally tree-like lattices~\cite{dhar2011hard, kundu2013reentrant, rodrigues2025hard}, but it is not clear that the high density phase on these graphs is similar to that on $d$-dimensional  hypercubic lattices.

Despite extensive work on dimer models, no exact solution is known for fully packed $k$-mers for any finite $k>2$ on a nontrivial lattice.  In this paper we provide the first such solution by determining the exact entropy of fully packed trimers on the kagome lattice.  The special geometrical structure of the kagome lattice allows a simple exact solution of the dimer model~\cite{misguich2003quantum, wang2007exact, 2008-dc-prl-exact}, and an analogous simplification occurs for trimers. We show that the fully packed trimer model on the kagome lattice is equivalent to a dimer model on the hexagonal lattice and to the ground states of the Ising antiferromagnet. Based on this, we show that the entropy of coverings of the kagome lattice per trimer is $ s_{\text{tri,kag}} = \frac{1}{2 \pi} \int_0^{ 2 \pi/3} \log( 2 + 2 \cos k) dk \approx 0.323065947\ldots$. In addition, we establish a two-to-one correspondence between trimer configurations and the time histories of walkers in a discrete-time totally asymmetric simple exclusion process (TASEP) on a line~\cite{kriecherbauer2010pedestrian}.

The remainder of the paper is as follows: In Sec.~\ref{sec:model}, we define the model and derive some simple lower bounds to the entropy. In Sec.~\ref{sec:transfer}, we describe the coordinate system and the characterization of configurations of trimers.  This is used to set up a transfer matrix for the problem. In Sec.~\ref{sec:tasep}, we establish a $2$-to-$1$ correspondence between trimer coverings and time histories of the evolution of a one-dimensional  discrete-time model  of diffusing hard core particles on a line. In Sec.~\ref{sec:equivalence}, we relate the counting of all possible histories of this model to the well-studied problem of dimers on a hexagonal lattice, which reduces to the calculation of ground state energy  of a free fermion gas, and thus determine the entropy of trimer coverings per site for the kagome lattice. Section~\ref{sec:summary} contains a summary, and some final remarks.

\section{Definition of the model \label{sec:model}}

Consider a kagome lattice with $3 N$ sites and periodic boundary conditions. The number of ways all the lattice sites can be fully covered by $N$ non-overlapping trimers varies as $\exp(s_{\text{tri,kag}} N)$, for large $N$. The entropy defined this way is dimensionless (we set the Boltzmann constant $k_B =1$).

\begin{figure}
\includegraphics[width=\columnwidth]{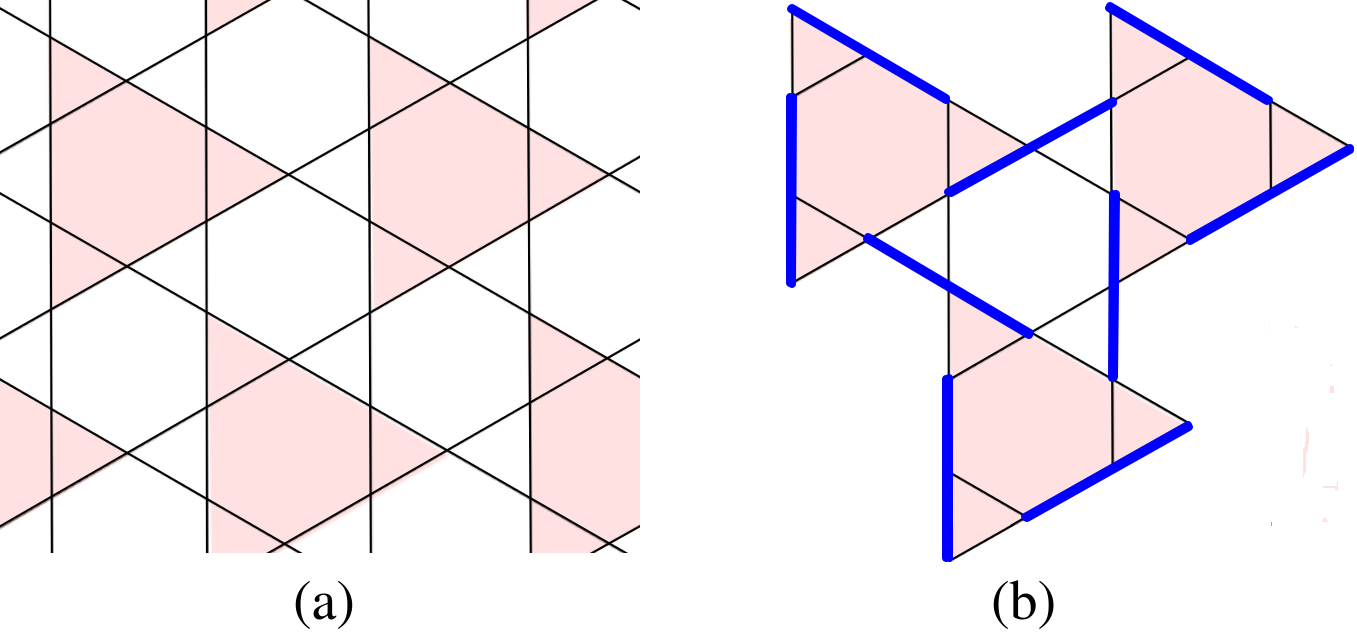}
\caption{(a) The kagome lattice as a union of non-overlapping equilateral triangles (pink colour) of side length $3$. Each triangle can be independently covered by trimers in two ways. (b) An example of covering of three adjacent pink triangles by trimers.}
\label{fig:lower_bound}
\end{figure}

It is easy to see that $s_{\text{tri,kag}}$ is non-zero. For instance, Fig.~\ref{fig:lower_bound}(a) shows that the vertices of the kagome lattice can be considered as a union of vertices of non-overlapping equilateral triangles of side length $3$, shown in pink colour. Each of these triangles has $9$ sites, and each can be covered by $3$ trimers in $2$ ways, independent of how other triangles are occupied. Thus the number of coverings of a torus made of such triangles having a total of $A$ sites has a lower bound $2^{A/9}$. Hence, we obtain
\begin{equation}
s_{\text{tri,kag}} \geq \frac{1}{3} \log 2 \simeq 0.23104906\ldots
\end{equation}

This lower bound can be improved by taking three adjacent triangles of the previous construction and making additional coverings of this set, where three of the trimers are shared between different triangles. There are $9$ ways to cover these three triangles by $9$ trimers, $8$ of them using the $2$ ways per triangle as before, and $1$ more, which is shown in Fig.~\ref{fig:lower_bound}(b). Thus the plane can be tiled fully with such triple-triangle tiles, leading to a better lower bound for the entropy of covering per trimer:
\begin{equation}
s_{\text{tri,kag}} \geq \frac{1}{9} \log 9 \simeq 0.24413606\ldots.
\end{equation}

\section{Setting up of the transfer matrix, and its block diagonal structure \label{sec:transfer}}

The number of coverings depends strongly on the boundary conditions imposed. In the case of domino tilings of aztec-diamond shaped regions, the phenomena of arctic circle has been studied~\cite{jockusch1998random}. In the following, we will make a particular choice: we consider the problem on a cylinder of finite width, which imposes periodicity in one direction (called the $x$-direction, see below). The cylinder will be taken to be semi-infinite in the $y$-direction, with some specified configuration of trimers in the bottom row and arbitrary configuration in the topmost occupied row. We will see that the entropy per site depends on the boundary condition at the bottom row, and we can determine the class of the configurations of this row for which the entropy per site is maximized. This choice of periodicity along one of the axes of the kagome lattice breaks its symmetry under lattice rotations, but makes the problem tractable.

\begin{figure}[b]
\includegraphics[width =\columnwidth]{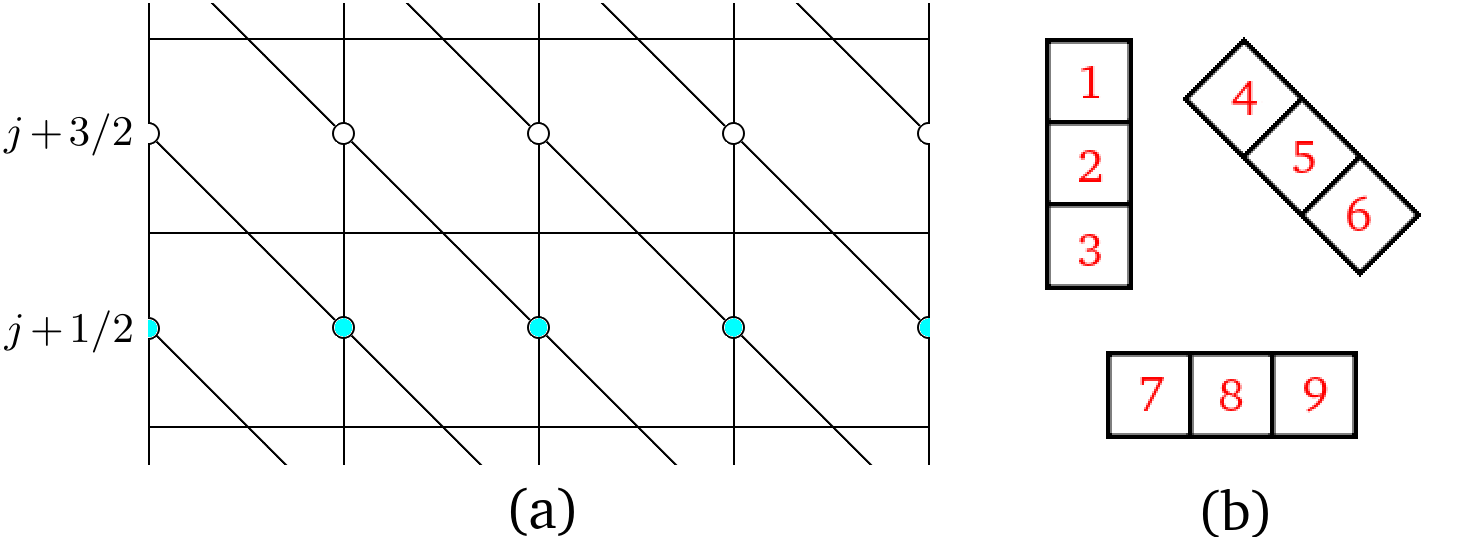}
\caption{(a) A kagome lattice strip (of width $L=4$) represented as a decorated square lattice. The two consecutive horizontal rows with the $y$ coordinates ($j+1/2)$ and $(j+3/2$) are marked by white and green circles, respectively. (b) To specify a covering, for each site, we specify the number, from $1$ to $9$, of the tile covering that site.}
\label{fig:lattice}
\end{figure}

\begin{figure*}
\includegraphics[width=1.9\columnwidth]{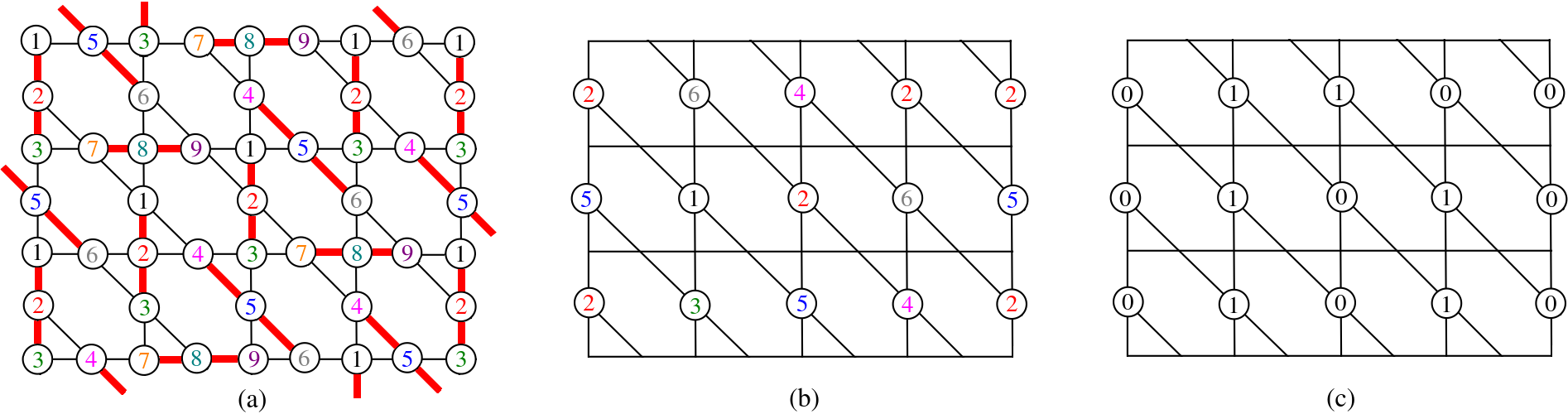}
\caption{(a) Example of a trimer covering of a part of the kagome lattice specifying the trimer state for each site. (b) Only states of sites in the horizontal rows constituting one sublattice are given. (c) Sites of the sublattice shown in (b) are now numbered only as $0$ or $1$, depending on whether they are covered by a midpoint of a trimer or not.}
\label{fig:labellings}
\end{figure*}

We will use the coordinate system as shown in Fig.~\ref{fig:lattice}(a). The sites of the lattice have coordinates $(i,j), (i+1/2,j), (i,j+1/2)$, with $i, j$ being integers. We will assume periodic boundary conditions along the $x$-direction, with $1 \leq x \leq L$. Along the $y$-direction, we will assume a semi-infinite lattice, for $ y \ge 0$. In all discussions and calculations that follows, we will use the rows of sites with half-integer $y$ coordinate (highlighted by circles or squares in the figures) as our reference. Therefore, we will be referring to such a rows when mentioning `horizontal rows'. We characterize the state of a site covering by an integer between $1$ and $9$, as defined in Fig. ~\ref{fig:lattice}(b).

Since there are at most $9$ states per site, the number of allowed configurations for a lattice of $3N$ sites is less than $9^{3N}$. Actually, the number is much smaller, as there are strong constraints for allowed assignments of states to sites. In fact, it is enough to specify the state of sites on any one of the three sublattices of the kagome lattice. We will choose this sublattice to be that corresponding to horizontal rows having a half-integer $y$-coordinate. This is a more efficient encoding of the configurations of trimers [see Figs. \ref{fig:labellings}(a) and (b)]. A somewhat similar characterization of tiling configurations was used by Gagunashvili and Priezzhev in the case of covering of a square lattice by $k$-mers, for a general $k$~\cite{gagunashvili1979close}.

The configuration of one of these horizontal rows is specified by giving the numbers of the tiles that covers each of the sites of that row, as defined in Fig.~\ref{fig:lattice}(b). For the sites in these horizontal rows, the allowed labels are only between $1$ and $6$. A configuration on a row of $L$ sites is given by specifying a state of these sites in a string; e.g., $26555542$ specifies the state a particular row of $8$ sites.

We construct a transfer matrix formulation of the problem as follows: We assume the state of the sites at $y =0$ and $y=1/2$ specified in some way. Then, for $ y \geq 1$, we cover all sites of the lattice by trimers up to the sites in the horizontal row $y = r+1/2$, with $r$ a non-negative integer. There may be some protruding sites occupied by these in the horizontal rows $y =r+1$, and $r+3/2$, but no trimer is placed that does not cover at least one site with the $y$-coordinate $\leq r+ 1/2$. We denote by $N_r(C)$ the number of ways we can fill up to the horizontal row $y=r+1/2$, where the configuration in the horizontal row $y = r+1/2$ is $C$. Then there is a recursion equation for these restricted partition functions given by
\begin{equation}
N_{r+1}(C) = \sum_{C'} T(C, C') N_r(C').
\end{equation}
This defines the transfer matrix $T(C,C')$. 
Note that the transfer matrix takes up from the configuration in the horizontal row $y =r+1/2$ to the row $y =r+3/2$, summing over all allowed configuration of the $2 L$ sites on the line $y =r+1$.

Nominally, as configuration $C$ on a horizontal row has $L$ sites, and each site can be in states $1$ through $6$, thus, the size of the matrix is $6^L \times 6^L$. The actual number of configurations needed is much less. For an overwhelmingly large majority of the $6^L$ strings, one cannot extend the configuration to higher values of $r$ without creating holes. Such vectors correspond to a zero eigenvalue of $T$, and may be ignored, without any loss. The actual number of configurations that need to be kept, $N_L$, is determined numerically for small values of $L$, it is given in Table I and grows approximately as $2.78897^L$.

\begin{figure}[t]
\includegraphics[width=8cm]{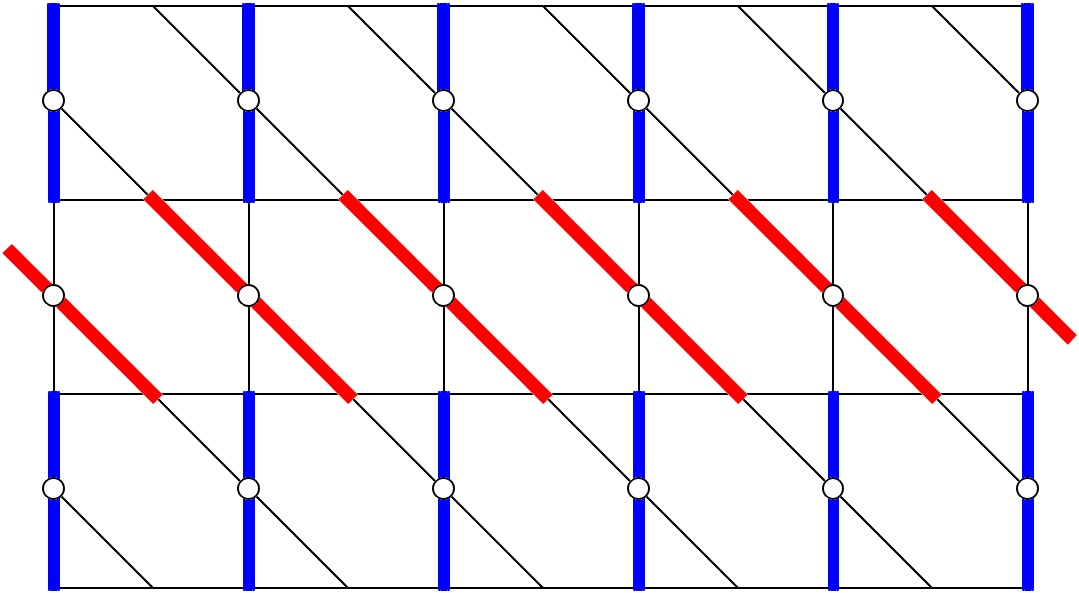}
\caption{The reference configuration for covering the kagome lattice by trimers.}
\label{fig:reference_conf}
\end{figure}

This transfer matrix can be simplified further, by noting that it has a block diagonal structure. This may be seen as follows:
If the configuration on a horizontal row is all $2$'s, then the only choice for the next layer is all $5$'s, and the next row above has to be again all $2$'s. Thus, these two configurations make a block of size $2$ of the transfer matrix. We will call the resulting two-dimensional configuration the reference configuration of trimers [Fig.~\ref{fig:reference_conf}]. There are actually two reference configurations, depending on whether the $j$-th horizontal row, with $j$ even, consists of all $2$'s, or all $5$'s. We will refer to these as $A$ and $B$ configurations.

Consider now a slightly more complicated case. Take $L = 8$, and consider $C' = 22355542$. Then it can be easily verified that the only allowed values of $C$ are $55122235$, $55122265$, $55122355$ and $55122655$. Now we try to construct possible allowed configurations $C''$ of the next higher horizontal row, and so on. We find that in all such configurations, there are only two sites with state not equal to $2$ or $5$, and the only allowed ones are of the type $ 2^s X 5^t Y$ (or those related to these by cyclic shifts), where $X= 3$ or $6$, and $ Y= 1$ or $4$, and $0 \leq s \leq 6$, and $t =6-s$. Together, all these configurations form a block of the transfer matrix for $L=8$.

Then we consider a configuration outside these blocks, and construct the configurations that are reachable from it, under the transfer matrix using the full coverage condition. These form another block. If this does not exhaust all configurations, we can start with a configuration not belonging to any of these blocks, and look at the set reachable from it, and so on, till we have exhausted all configurations. If the configuration does not allow a full covering extension, it is discarded. This constructs the full transfer matrix.

For any configuration $C$, let $n_u(C), n_m(C), n_l(C)$ denote the number of sites covered by the top end, middle, and bottom end of the covering trimer. We find that for all configurations $C$ within a block, the values of $n_u(C), n_m(C), n_l(C)$ are the same. In addition, $n_u(C)=n_l(C) = (L- n_m(C))/2$ . Thus, each block may be characterized by a single even integer $n_u +n_l$.

In fact, we find that for all non-trivial configurations that allow a full covering of the lattice, a much more reduced description of $C$ is sufficient. We denote the state of a site by a $0$, if the site is in the state $2$ or $5$, and by $1$ if it is any other state. Then the configuration is represented by an $L$-bit binary vector. For example, a configuration $235512$, has the binary representation $010010$ [see Figs. \ref{fig:labellings}(b) and (c)].

We may think of the $0$'s as empty sites, and the $1$'s as occupied by hard-core particles. Corresponding to each valid covering of the lattice by trimers, we get a configuration of $0$'s and $1$'s on the horizontal rows. By thinking of the configuration on the row $y =r+1/2$ as particles on a line at time $r$, then, each lattice configuration of trimers encodes a time history of the evolution of these particles. It turns out that this time evolution is particularly simple. The number of particles is conserved in time, and each particle moves in a discrete time, in the well-studied totally asymmetric simple exclusion process (TASEP)~\cite{golinelli2006asymmetric}. A particle at time $t$ can either stay at the same site, or jump one unit to the left. If the jump will result in two particles at the same site, the move is not allowed. All sites are updated in parallel. If there are more than one particle occupying consecutive sites on the line, all, or a subset of them can move one unit to the left together, if that does not violate the hard-core constraint. Also, for TASEP histories corresponding to trimer configurations, the total number of particles is always even.

Clearly, for each trimer configuration, the binary representation gives a unique space-time trajectory of the TASEP evolution.

\section{ The $2$-to-$1$ correspondence between trimer configurations and TASEP histories \label{sec:tasep}}

We will now argue that the correspondence between the trimer configurations, and TASEP histories is two-to-one: For each TASEP history, there are exactly two trimer configurations.

As mentioned above, the number of particles is conserved in the TASEP process. Where there are no particles, there is a single TASEP history, and this corresponds to the reference configurations A and B.

\begin{figure}
\includegraphics[width=8cm]{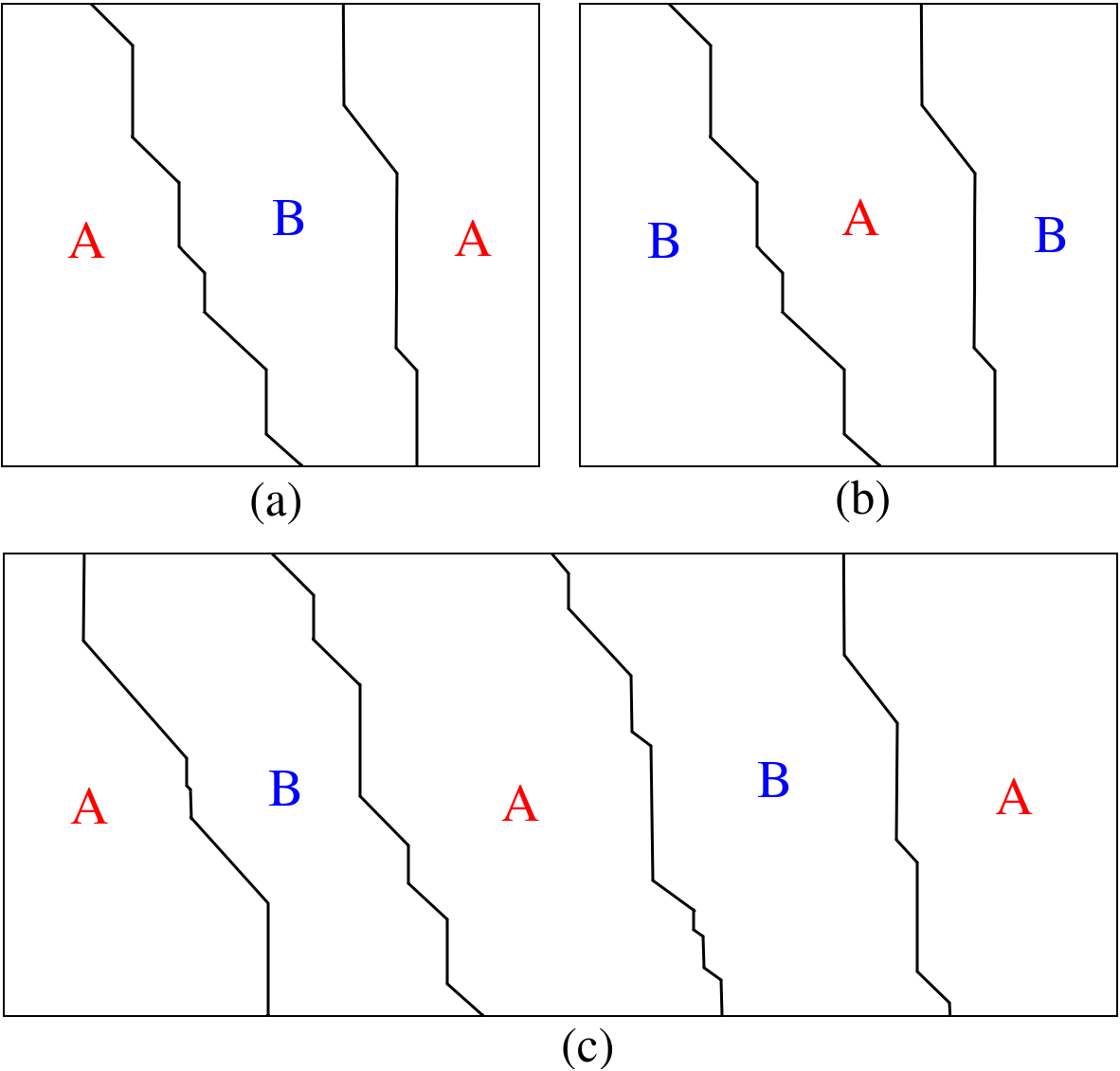}
\caption{Schematic representations of three trimer configurations. In (a) and (b), they correspond to the same TASEP configuration, having only two particles. The world-lines of the particles are boundaries between A- and B-type regions; and we get two different allowed trimer configurations, if we choose the regions to be A-B-A or B-A-B. Panel (c) presents a TASEP history with four particles.}
\label{fig:many_lines}
\end{figure}

Consider now a two-dimensional space-time history with two walkers, represented schematically in Figs. \ref{fig:many_lines}(a) and (b). The world lines of walkers separate the regions of space-time that are A and B regions in the trimer configurations. If we assign a state $2$ or $5$ to one of the sites in one such connected region, the state of all other sites in the region is fixed (same in the same horizontal row, and $2$ and $5$ alternate on adjacent rows). Since on crossing a defect-line, one makes a transition from an A region to a B region (and vice versa), this also fixes the state of sites in the other regions.

We can also determine the state of sites on the world-lines of the walkers (boundaries between regions). For instance, if the particle has state $5$ on the left, and state $2$ on the right, then: It is assigned a state $1$, if the particle at the previous time step was at the same $x$-coordinate; and a state $4$, if it was at $(x+1)$. If the site has left neighbor $2$ and right neighbor $5$, then: It is assigned a state $3$ if the world-line stays at the same site at the next time; and the state $6$ otherwise. If two particles are adjacent, one imagines a virtual empty region of opposite type separating them. Thus, the local TASEP configuration $0110$, will be treated as $01(0)10$, where the $(0)$ indicates a fictitious zero space to keep track of the parity of the regions (starting in the A-region, and crossing two boundaries, we again find an A-region).

If there are more than two particles in the TASEP, or more than two `defect boundaries' in the trimer configurations, these local rules do not change.

In Fig.~\ref{fig:many_lines}(a) and (b), we show a local region of a larger lattice with full coverings of the lattice by trimers, where the left region (having A-ordering) is separated from the right region (with type-B ordering) by a line going from top to bottom. This boundary line is of width 1. Of course, there are also defect lines with type-B region on the left. In Fig.~\ref{fig:many_lines}(b), we show such a configuration, where the defect line has the same position as in (a), but the type-A region is now to the right.

In Fig.~\ref{fig:many_lines}(c), we show a schematic representation of a more complicated situation with many such boundary lines. We will also refer to these as defect lines. Since the constraints of full coverage are local constraints, it is clear that one can construct the corresponding trimer configuration in these cases. Moreover, due to the periodic boundary conditions, the total number of defect lines has to be even (let us say, equal $2m$). Once again, each line at next row can either stay at same position $x$, or move one unity to the left. Note that if there are $r$ lines in consecutive positions at a given row, each of them can take a step to the left without violating the no-crossing condition. These paths are thus one-to-one correspondence with those of $2m$ particles (or walkers) in the TASEP. We would like to emphasize that this correspondence is between the allowed configurations in the two cases, and we use this correspondence to set up the transfer matrix. The weights of the configurations in the two problems are quite different.

Specifying the motion of the $2m$ walkers specifies the full trimers configuration essentially uniquely. To be precise, there are exactly two fully-packed trimers configurations for each valid space-time history of the walkers, corresponding to the freedom to choose the state of the first region as A or B. For example, for $L=10$, we can have a TASEP configuration $0100110100$, and the corresponding trimer configurations on the row are $2x55yx5y22$ or $5y22xy2x55$, where $x$ can be $3$ or $6$, and $y$ can be $1$ or $4$. Note that sequences of $2$'s and $5$'s alternate, and a sequence of $2$'s ($5$'s) starts with a $y$ ($x$) and ends with an $x$ ($y$). For this reason, $x$ and $y$ denote the boundary sites. Specification of the TASEP configuration at one time does not uniquely fix where the site marked as $y$ is $1$ or $4$, for example. Which one of these choices is fixed by whether the TASEP particle stays put or moves one unit to the left.

\begin{figure}
\includegraphics[width=8.5cm]{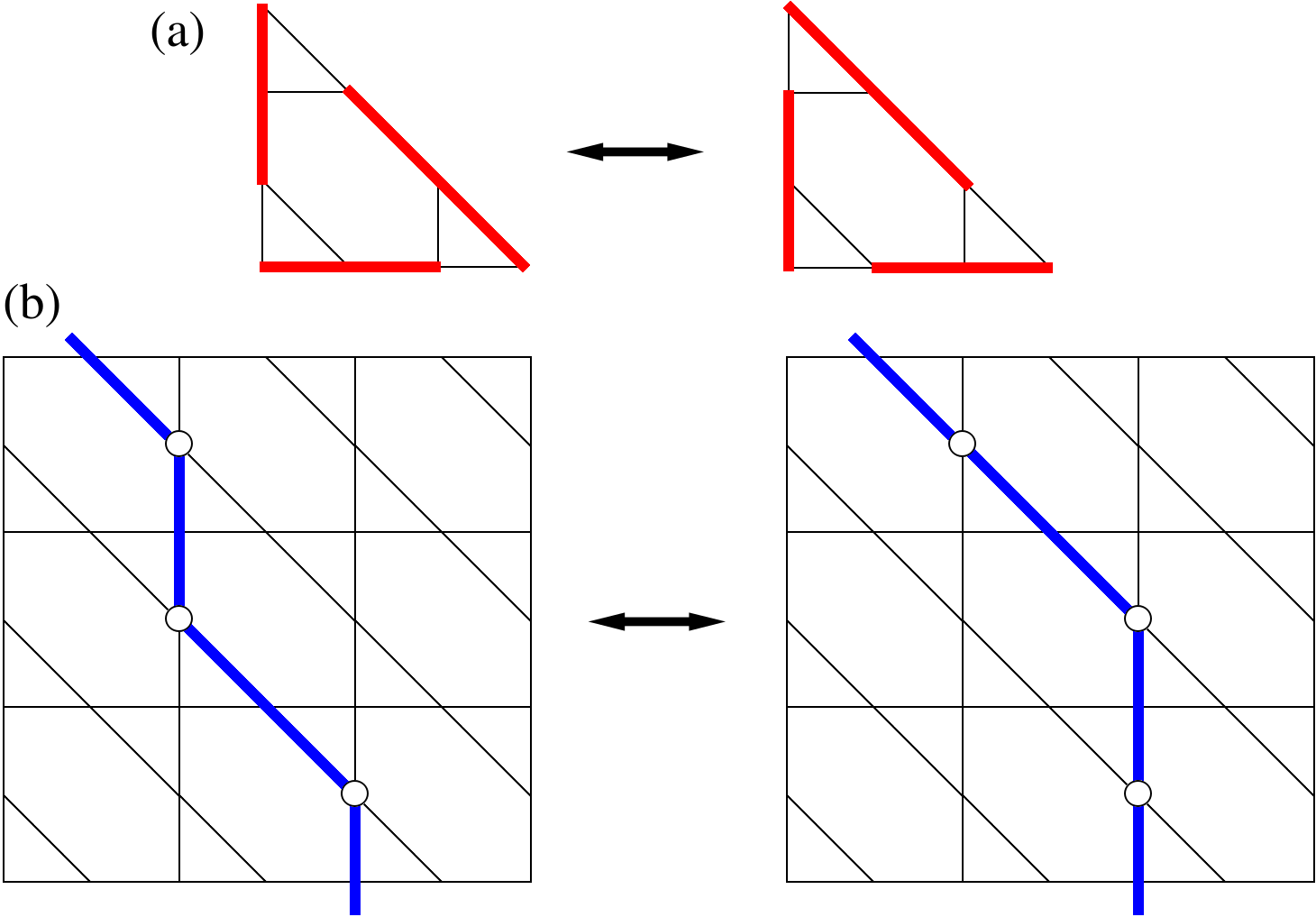}
\caption{The basic flip move for (a) trimer coverings and (b) TASEP paths.}
\label{fig:flip-move}
\end{figure}

There exists an assignment of configuration of trimers for every valid history $\mathcal{B}$ of the walkers. Figures~\ref{fig:flip-move}(a) and (b) show that if we change a time-history by changing a walk locally from `up-left' to `left-up', the corresponding trimer configuration is also changed locally, by a `triangle move', involving rearranging only three trimers. But we can go from any history $\mathcal{B}$ to any other history by a sequence of such moves, and hence for any other history $\mathcal{B}'$, also, there is a unique configuration of trimers that can be obtained by such local moves.

\section{ Equivalence to dimer coverings of the hexagonal lattice and determination of the eigenvalues of the transfer matrix \label{sec:equivalence}}

\begin{figure*}
\includegraphics[ width=2\columnwidth]{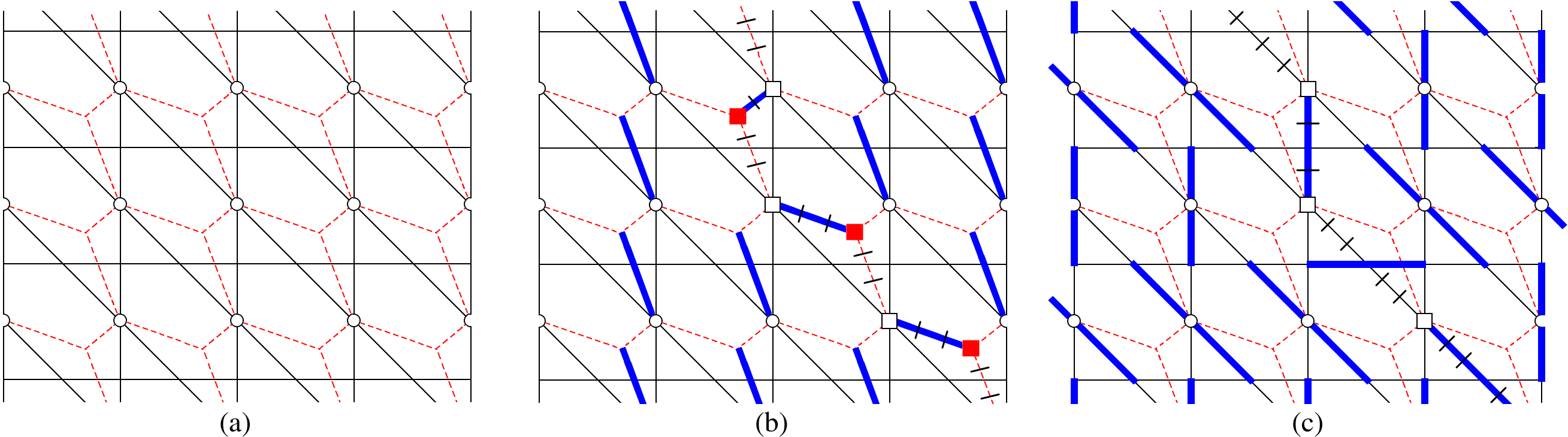}
\caption{(a) The kagome lattice (solid) and the corresponding hexagonal lattice (dashed lines). The sites belonging to the horizontal rows of the kagome lattice, shown as open circles, are common to the hexagonal lattice also. Panel (b) presents a configuration of dimers placed on the hexagonal lattice, and one of the corresponding trimer configurations on the kagome lattice is depicted in (c). Dimers and trimers are represented by blue bonds, the defect lines are marked by ticks, and the boundary sites are denoted by squares. Note that the position of the boundary sites on the kagome lattice (denoted by open squares) is the same in both (b) and (c) panels.}
\label{fig:hexag}
\end{figure*}

It is easy to see that there is an equivalence of the above described TASEP walkers to the model of fully-packed dimers on the hexagonal lattice. The difference graph between any fully packed configurations of dimers and a standard configuration of dimers consists of non-intersecting lines (called defect lines here) that may be considered as time histories of hard core biassed random walkers \cite{damle2012resonating}. We use this correspondence here to set up a two-to-one correspondence between the configurations of the trimers on the kagome lattice to dimers on the hexagonal lattice. We start drawing the hexagonal lattice as shown in Fig.~\ref{fig:hexag}(a), where half of its $2 L^2$ sites are shared with the kagome lattice (open circles) and the other half are not. By considering the covering of the hexagonal lattice by dimers, we may define the reference configuration as the one where all the bonds have a kagome lattice site as their lower site. Then, it is easy to see that the defect lines in the dimer model are the same as the TASEP walkers in the trimer model; compare Figs.~\ref{fig:hexag}(b) and \ref{fig:hexag}(c).

We also mention another well-known relation of the configurations of full covering of the hexagonal lattice by dimers to the ground-states of the nearest neighbor antiferromagnetic Ising model on the dual triangular lattice~\cite{dhar2000triangular}.

It is straightforward to generalize this treatment to the case of three different activities $z_1, z_2, z_3$ for trimers with the three different orientations. In this case, we define the partition function as the sum of weights of all coverings, where the weight of a configuration having $n_1, n_2, n_3 $ trimers with orientations in the three possible directions is $z_1^{n_1} z_2^{n_2} z_3^{n_3}$. A fully ordered configurations with no defects has a weight $z_1^N z_2^N$ for a fully covered lattice with $6N$ sites. Taking out a normalization  factor $(z_1 z_2 z_3)^N$, we can equivalently assign a weight $1/z_1$ to step on a defect path that goes straight, $1/z_2$ to a trimer that goes left, and $1/z_3$ to a trimer not on the defect path. The relative weights of configurations then become equivalent to a dimer model on the hexagonal lattice with activities $(1/z_1, 1/z_2,1/z_3)$. We note that if $1/z_1 > (1/z_2 + 1/z_3)$, the state is a fully ordered layered state with alternating $222..$ and $555...$ horizontal rows, and the density of defect lines can be made small by making $(1/z_2+1/z_3 - 1/z_1)$ small. Thus, it is useful to consider the number of walkers =$0,2,4 \ldots$ as a parameter we can vary.

Consider the block of the transfer matrix corresponding to the TASEP configuration with $2m$ particles. If the configuration $C'$ has defects at positions $R_1,R_2, ..,R_{2m} \equiv \{R_j\}$, then the eigenvectors $\Psi(\{R_j\}) $ satisfy an equation of the form
\begin{equation}
T \Psi( \{R_j\}) = \sum_{\{R_j\}'} T( \{R_j\},\{R_j\}') \Psi(\{R_j\}').
\end{equation}

In this approach, the problem becomes equivalent to an earlier studied problem related to the ground state of the antiferromagnetic Ising model on the triangular lattice \cite{dhar2000triangular, stephan2021extreme}. Since the details of this calculation are identical to what is discussed in \cite{dhar2000triangular}, we will skip the details here, and only mention the result. The problem is found to be equivalent to a problem of free fermions. For $2m$ defect lines in a strip of width $L$, the eigenvalues of the TM are given by
\begin{equation}
\lambda_L = \prod_{\ell=1}^{2m} [ 1 + \exp( i k_{\ell})],
\label{eq:FFevs}
\end{equation}
where $ k_{\ell}$ are distinct numbers satisfying $ \exp(i k_{\ell} L) = -1$. The ${L \choose 2m}$ different choices of selecting
$2m$ values out of the $L$ roots of this equation give the different eigenvalues of the TM. The corresponding eigenfunction $\Psi(\{R_j\})$ is a Slater determinant with $2m$ orthogonal single particle states $\phi_{\ell}(x) = \exp(ik_{\ell} x)$.

To obtain the largest eigenvalue, we occupy all single-particle states with momentum $k$ with $ |1 + e^{ik}| > 1$.
In the limit of large $L$, the allowed values of $k$ are $ -2 \pi/3 \leq k \leq 2 \pi/3$. And we find
\begin{equation}
\frac{\log \lambda_L^{(max)}}{L} = \frac{1}{2 \pi} \int_{-2 \pi/3}^{2 \pi/3} \log | 1 + e^{ik}| dk.
\end{equation}
Thus, the entropy \textit{per trimer} for the covering of the kagome lattice by trimers is
\begin{equation}
s_{\text{tri,kag}} = \frac{1}{2 \pi} \int_0^{ 2 \pi/3} \log( 2 + 2 \cos k) dk. \end{equation}
This is the same result as the entropy per dimer $s_{\text{dim,hex}}$ for the dimer covering of the hexagonal lattice, as given in Ref. \cite{wannier1950antiferromagnetism} (though there is a typo in the numerical result reported there). Therefore,
\begin{equation}
s_{\text{tri,kag}} = s_{\text{dim,hex}} = 0.323065947\ldots
\label{eq:entropy}
\end{equation}
The entropies \textit{per site} (i.e., $s_{\text{tri,kag}}/3$ and $s_{\text{dim,hex}}/2$) are obviously different in the two cases.

\begin{table}
\caption{Dimension of the trimer transfer matrix ($N_L$), the number of their non-null eigenvalues ($N_{ev}$) and the finite-size values of the entropy ($s_L$) as functions of the strip width $L$.}
\begin{ruledtabular}
\begin{tabular}{rrrrr}
% \hline\hline
$L$ &   $N_L$  & $N_{ev}$ &   $s_{L}$        \\
\hline
2  &       10  &     3    &   0.346573590    \\
3  &       26  &     6    &   0.366204096    \\
4  &       58  &    12    &   0.306986794    \\
5  &      172  &    24    &   0.321887582    \\
6  &      478  &    48    &   0.335017512    \\
7  &     1304  &    96    &   0.318510444    \\
8  &     3674  &   192    &   0.321929761    \\
9  &    10214  &   384    &   0.328522389    \\
10 &    28460  &   768    &   0.321015269    \\
11 &    79444  &  1536    &   0.322290453    \\
12 &   221470  &  3072    &   0.326167962    \\
13 &   617684  &  6144    &   0.321917773    \\
14 &  1722808  & 12288    &   0.322524314    \\
15 &  4804636  & 24576    &   0.325061595    \\
16 & 13400154  & 49152    &   0.322336670    \\
%  \hline\hline
\end{tabular}
\label{tab:Rowstates}
\end{ruledtabular}
\end{table}

We have confirmed these results by explicitly calculating the TMs, considering all allowed row configurations (with site states in $\{1,\ldots, 6\}$), for strip widths $L= 2,3,\ldots,16$. Here, the TM size is much bigger than in the free fermion case (see Tab. \ref{tab:Rowstates}), being given by $N_L \simeq 2.78897^L$ for large $L$. As discussed above, the number of defect sites in a given row is preserved as one moves to the next row. By taking advantage of this, we have generated each block separately, avoiding the difficulties in obtaining and dealing with very large matrices. From the numerical diagonalization of such blocks, the full spectra of eigenvalues of these TMs were obtained. The largest eigenvalue is located in the block related to $2\lfloor (L+1)/3\rfloor$ defect lines. Interestingly, the number $N_{ev}$ of \textit{non-null} eigenvalues is given by $N_{ev}=3\times2^{L-2}$ (see Tab. \ref{tab:Rowstates}), which is much smaller than $N_L$. Indeed, a large portion of the eigenvalues of the trimer TM is null. On the other hand, $N_{ev}$ is larger than the corresponding number in the free fermion case, which is $2^{L-1}-1$.

We verified that all eigenvalues obtained from Eq. \ref{eq:FFevs}, for given $L$ and $m$, are contained in the largest set obtained for the corresponding block of the TM of trimers. Importantly, the dominant eigenvalue in each block is always the same in both cases, demonstrating that the higher number of basis vectors in the trimer case does not affect the calculated value of the entropy. We do not have a general proof, but we have verified that this procedure is clearly working for number of walkers $ m=0,2,4 \ldots$ for all the $L$ values we have studied. The numerical values for the entropies obtained in this way are displayed in Tab. \ref{tab:Rowstates}. Their convergence to the asymptotic theoretical result is approximately as $1/L^2$, with oscillatory corrections of period 3. Notably, the relative difference between the exact and the numerical result for $L=16$ is only $0.2$\%.

\section{summary and concluding remarks  \label{sec:summary}}

In this paper, we showed that there is a two-to-one equivalence between the configurations of full coverings of the kagome lattice by trimers, and the full covering of the corresponding hexagonal lattice by dimers. Then, the exact value of the entropy per trimer in the first problem is exactly the same as the entropy per dimer in the second problem, in the thermodynamic limit.

\begin{figure}
\includegraphics[width=8.5cm]{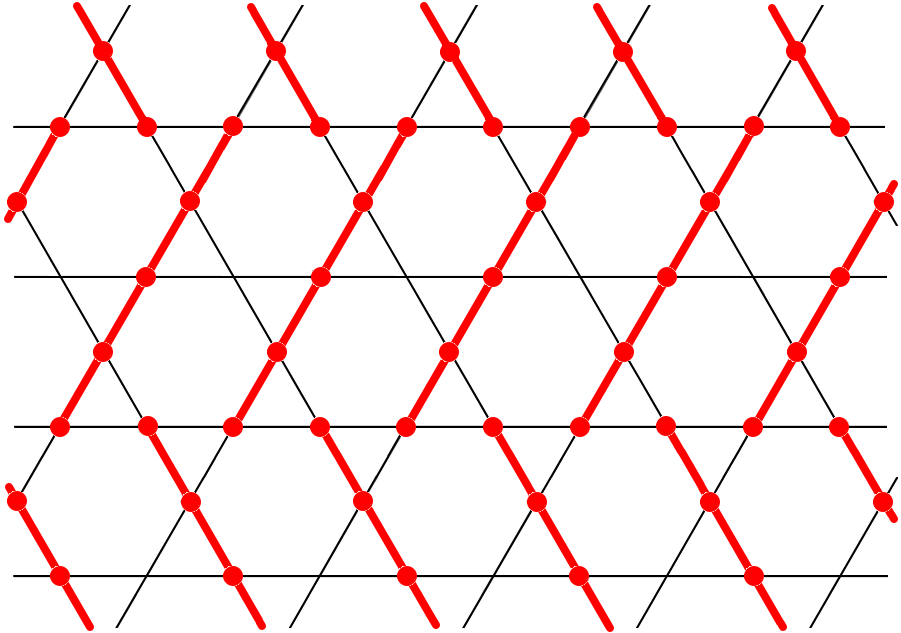}
\caption{Example of a close-packed configuration of 5-mers on a kagome lattice strip of width $L=5$.}
\label{fig:closepackk5}
\end{figure}

The exact value of the close-packing density for $k$-mers on the kagome lattice can be shown to be $\rho_k \geq \frac{2 k}{ 3 (k-1)}$, for $k \geq 3$. For instance, Fig. \ref{fig:closepackk5} shows an example of a close-packed configuration for $k=5$, where one sees that all horizontal rows with $L$ sites are fully occupied, while the rows with $2L$ sites alternate between fully- and half-occupancy, yielding $\rho_5 = 5/6$. It will be interesting to determine if this expression for $\rho_k$ is actually exact for $k >3$, and whether the entropy at the closest packing is zero or nonzero.

We noted that in the transfer matrix formulation, if one naively constructs the transfer matrix of the problem for a cylinder of width $L$, it will be of dimension $6^L \times 6^L$. However, most of the corresponding basis vectors in the configuration basis have zero eigenvalue. Given a row configuration $C$, there is no satisfactory algorithm to decide if it will allow the configuration to be continued to higher rows or not without creating holes. We have not been able to formulate an algorithm that can optimally reduce the size of the transfer matrix for general $L$. For higher $k >3$, the question is more complicated, as the maximum packing density is less than one, and the existence of a single vacant sites is no longer a valid test of non-maximality of the covering.

The number of basis vectors in the TASEP formulation is much smaller. This is partly due to the fact that a single basis vector in the TASEP formulation corresponds to a linear combination of several trimer basis vectors. For example, the $L=3$ basis vector for the TASEP $ |110\rangle$ in the occupation number basis, corresponds to the linear combination of eight trimer basis vectors
$$|135\rangle +|165\rangle + |312\rangle + |342\rangle + |435\rangle+ |465\rangle+ |612 \rangle + |642\rangle.$$
Also, these vectors are not always with equal weights. There is also no simple formula to express the relationship between these basis vectors.

The dimer problems are a topic of much current interest, in both classical~\cite{flicker2020classical, lloyd2022statistical, ramirez2024dimer, baker2025entropies} and quantum \cite{satzinger2021realizing, semeghini2021probing, yan2022triangular, singh2024exact, zeng2025quantum} fronts. Also, there is a lot of technological interest in kagome-like materials, due to they remarkable physical properties~\cite{yin2022topological}. Their production (e.g., via self-assemble) on the surface of different materials have been recently reported \cite{yan2021synthesis, yan2021two, kumar2021manifestation, lin2022fabrication, farinacci2024yu}. It may be hoped that our work would also help to understand these materials and processes.

%\section{Acknowledgments}
\acknowledgments
DD's work is supported by the grant SP/DP/2023/658 by Indian National Science Academy. TJO acknowledges partial support from CNPq and Fapemig (Brazilian agencies). JFS is partially supported by CNPq.

%\bibliography{ref}

%merlin.mbs apsrev4-1.bst 2010-07-25 4.21a (PWD, AO, DPC) hacked
%Control: key (0)
%Control: author (8) initials jnrlst
%Control: editor formatted (1) identically to author
%Control: production of article title (-1) disabled
%Control: page (0) single
%Control: year (1) truncated
%Control: production of eprint (0) enabled
%

\end{document}